\documentclass{PoS}

\usepackage{epsfig}

\PoS{PoS(LAT2005)339}

\title{$B_7$, $B_8$ and chiral Ward identities }

\ShortTitle{$B_7$, $B_8$ and chiral Ward identities }

\author{\speaker{Weonjong Lee} \thanks{This research is supported in
part by the KOSEF grant (R01-2003-000-10229-0), by the MOST/KISTEP
grant of international collaboration, and by the BK21 program.}\\
Center for Theoretical Physics, Department of Physics, Seoul National
University, Seoul, 151-747, South Korea \\ E-mail:
\email{wlee@phya.snu.ac.kr}}

\author{George T. Fleming\\
Sloane Physics Laboratory, Yale University, New Haven, CT 06520, USA\\
E-mail: \email{George.Fleming@Yale.edu}}

\abstract{ We present recent progress in understanding weak matrix
elements on the lattice. We use HYP staggered fermions in quenched QCD
to study numerically various properties of the $K^+\rightarrow\pi^+$
amplitudes of the electroweak penguin operators $Q_7$ and $Q_8$. We
check chiral Ward identities to probe the validity of using improved
staggered fermions in the calculation of weak matrix elements. We
address the issue of mixing with unphysical lower dimension operators,
which causes a divergent term in the case of the $\Delta I = 1/2$
amplitudes. We propose a particular subtraction method as the best
choice. We also measure the gold-plated ratio $R$ originally suggested
by Becirevic and Villadoro.}

\FullConference{XXIIIrd International Symposium on Lattice Field Theory\\
25-30 July 2005\\
Trinity College, Dublin, Ireland}

\begin{document}
\section{Introduction to Chiral Ward Identities}
\label{sec:WI}
By decoupling the heavy particles such as W bosons, Z bosons and heavy
quarks from the standard model of electroweak theory, we obtain the
low energy effective Hamiltonian relevant to particular processes
classified as the $\Delta S =1$ interaction of primary interest in
this paper.
\begin{eqnarray}
{\cal H}^{\Delta S = 1} = \frac{G_F}{\sqrt{2}} V_{ud} V^{*}_{us}
\sum_{i=1}^{10} [ z_i(\mu) + \tau y_i(\mu) ] Q_i(\mu)
\end{eqnarray}
Here, $G_F$ is the Fermi coupling constant and $V_{ij}$ are elements
of the CKM matrix.
Note that $\tau = - \lambda_t / \lambda_u$ 
where we define $\lambda_i \equiv V_{id} V^*_{is}$.
The $z_i(\mu)$ and $y_i(\mu)$ are the Wilson coefficients at the scale
$\mu$.
The $Q_i(\mu)$ are the four fermion operators made of $u$, $d$, $s$
quark fields.
The Wilson coefficients for the current-current operators $Q_1$ and
$Q_2$ are of ${\cal O}(1)$, those for the QCD penguin operators
($Q_i$, $i=3,4,5,6$) are of ${\cal O}(\alpha_s)$, whereas those for
the electroweak penguin operators ($Q_i$, $i=7,8,9,10$) are of ${\cal
O}(\alpha)$.
In this paper, we focus on the electroweak penguin operators $Q_7$ and
$Q_8$.
\begin{eqnarray}
Q_7 &=&  \frac{3}{2} ( \bar{s}_\alpha d_\alpha)_{V-A}
\sum_{q=u,d,s} e_q ( \bar{q}_\beta q_\beta )_{V+A}
\\
Q_8 &=& \frac{3}{2} ( \bar{s}_\alpha d_\beta)_{V-A}
\sum_{q=u,d,s} e_q ( \bar{q}_\beta q_\alpha )_{V+A}
\end{eqnarray}
According to the group theory analysis, $Q_7$ and $Q_8$ belong to the
$(8,8)$ irreducible representation (irrep) of $SU(3)_L \otimes SU(3)_R$.
We can further decompose $Q_{(8,8)}$ into $\Delta I = 1/2$ and
$\Delta I = 3/2$ irrep.
\begin{eqnarray}
Q_{(8,8)} &=&  \frac{1}{2} ( \bar{s} d)_{V-A}
( 2 \bar{u} u - \bar{d}{d} - \bar{s}{s} )_{V+A}
\label{eq:Q(8,8)}
\\
Q_{(8,8)}^{\Delta I=3/2} &=& (\bar{s} d)_{V-A} (\bar{u} u)_{V+A}
+ (\bar{s} u)_{V-A} (\bar{u} d)_{V+A}
- (\bar{s} d)_{V-A} (\bar{d} d)_{V+A}
\label{eq:Q(8,8) 3/2}
\\
Q_{(8,8)}^{\Delta I=1/2} &=& (\bar{s} d)_{V-A} (\bar{u} u)_{V+A}
- (\bar{s} u)_{V-A} (\bar{u} d)_{V+A}
- (\bar{s} d)_{V-A} (\bar{s} s)_{V+A}
\label{eq:Q(8,8) 1/2}
\end{eqnarray}
We can rewrite $Q_{(8,8)}$ as follows:
\begin{eqnarray}
Q_{(8,8)} &=&
\frac{1}{2} [ Q_{(8,8)}^{\Delta I=3/2} + Q_{(8,8)}^{\Delta I=1/2} ]
\end{eqnarray}

Using chiral perturbation theory, we can analyze the chiral
behavior of hadronic matrix elements of the $Q_{(8,8)}$ operators.
The details of this analysis will be presented in
Ref.~\cite{ref:wlee:1}.
At the lowest order in chiral perturbation theory, the ratio of the
$K^+ \rightarrow \pi^+$ amplitudes satisfies the following chiral Ward
identity \cite{ref:rbc:1,ref:laiho:1,ref:wlee:1}.
\begin{eqnarray}
& & W \equiv \frac{ \langle \pi^+ | Q_{(8,8)}^{\Delta I=1/2} | K^+
\rangle_{\rm sub} } { \langle \pi^+ | Q_{(8,8)}^{\Delta I=3/2} | K^+
\rangle },
\qquad
\lim_{m_q \rightarrow 0} W = 2
\label{eq:WI:1}
\end{eqnarray}
In this paper, one of the main goals is to numerically check the above
Ward identity, which serves as an important probe to test the validity
of using improved staggered fermions in the calculation of weak matrix
elements.
\begin{table}[h!]
\begin{center}
\begin{tabular}{c | c}
\hline
parameter & value \\
\hline
gauge action & Wilson Plaquette \\
fermion action & HYP Staggered \\
$\beta $ & 6.0 (quenched QCD) \\
\# of confs & 218 \\
lattice & $16^3 \times 64$ \\
quark mass & 0.01, 0.02, 0.03, 0.04 \\
\hline
\end{tabular}
\end{center}
\caption{Parameters for the numerical study}
\label{tab:param}
\end{table}
\section{Numerical Study on Ward identities}
\label{sec:num}
We use an ensemble of 218 gauge configurations whose details
are summarized in Tab.~\ref{tab:param}.
We use the $\rho$ meson to set the scale $1/a=1.95$ GeV.
We also use the kaon to set the physical strange quark mass.
Details of this procedure are presented in Ref.~\cite{ref:wlee:2}.
Throughout this paper, we consider only particles composed of
degenerate quarks ($m_s = m_d$).
One of the key ingredients of this numerical study is that we use
improved staggered fermions made of HYP fat links when we construct
the operators and calculate the matrix elements.
\begin{figure}[h!]
\epsfig{file=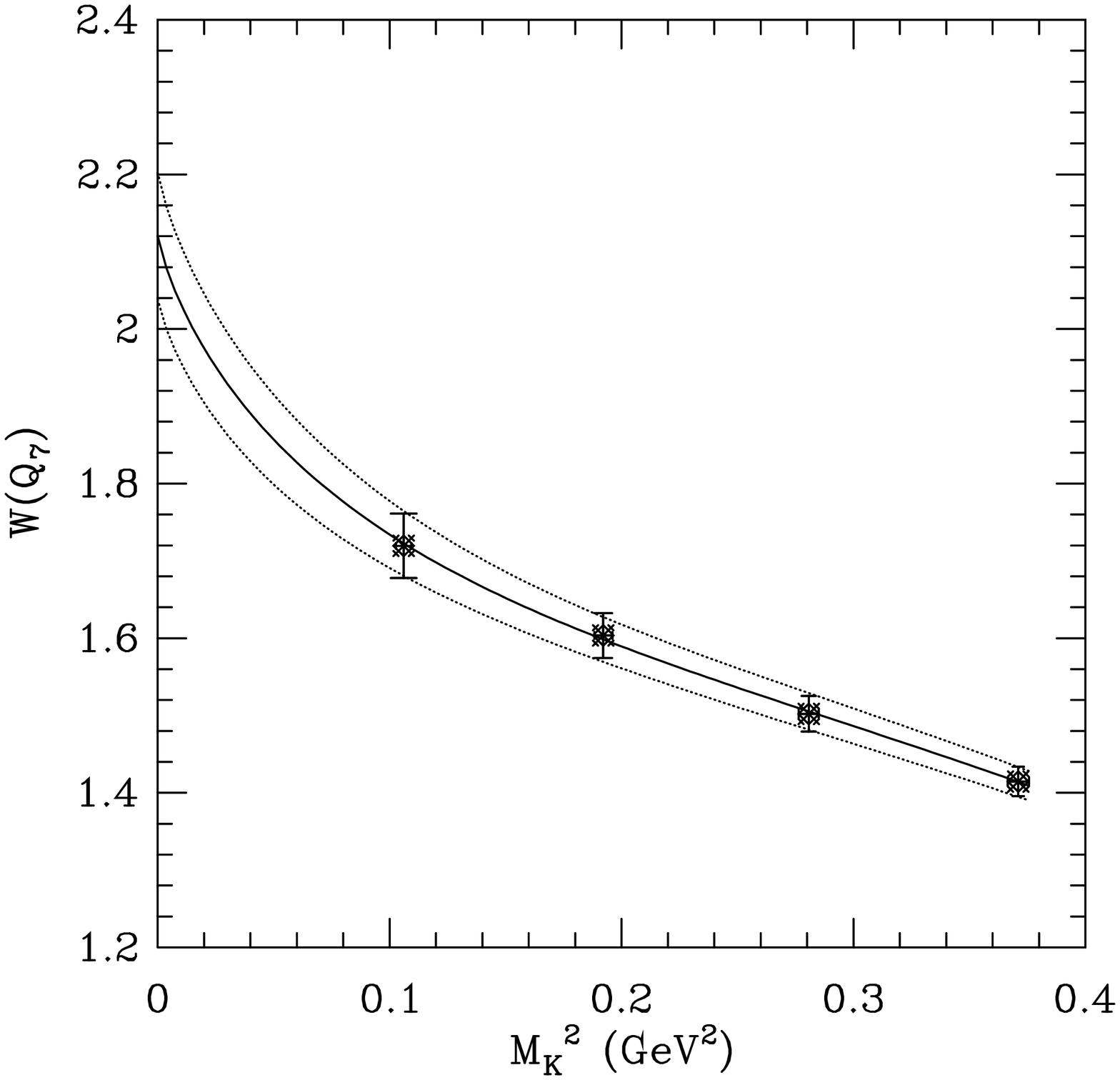, width=0.5\textwidth}
\epsfig{file=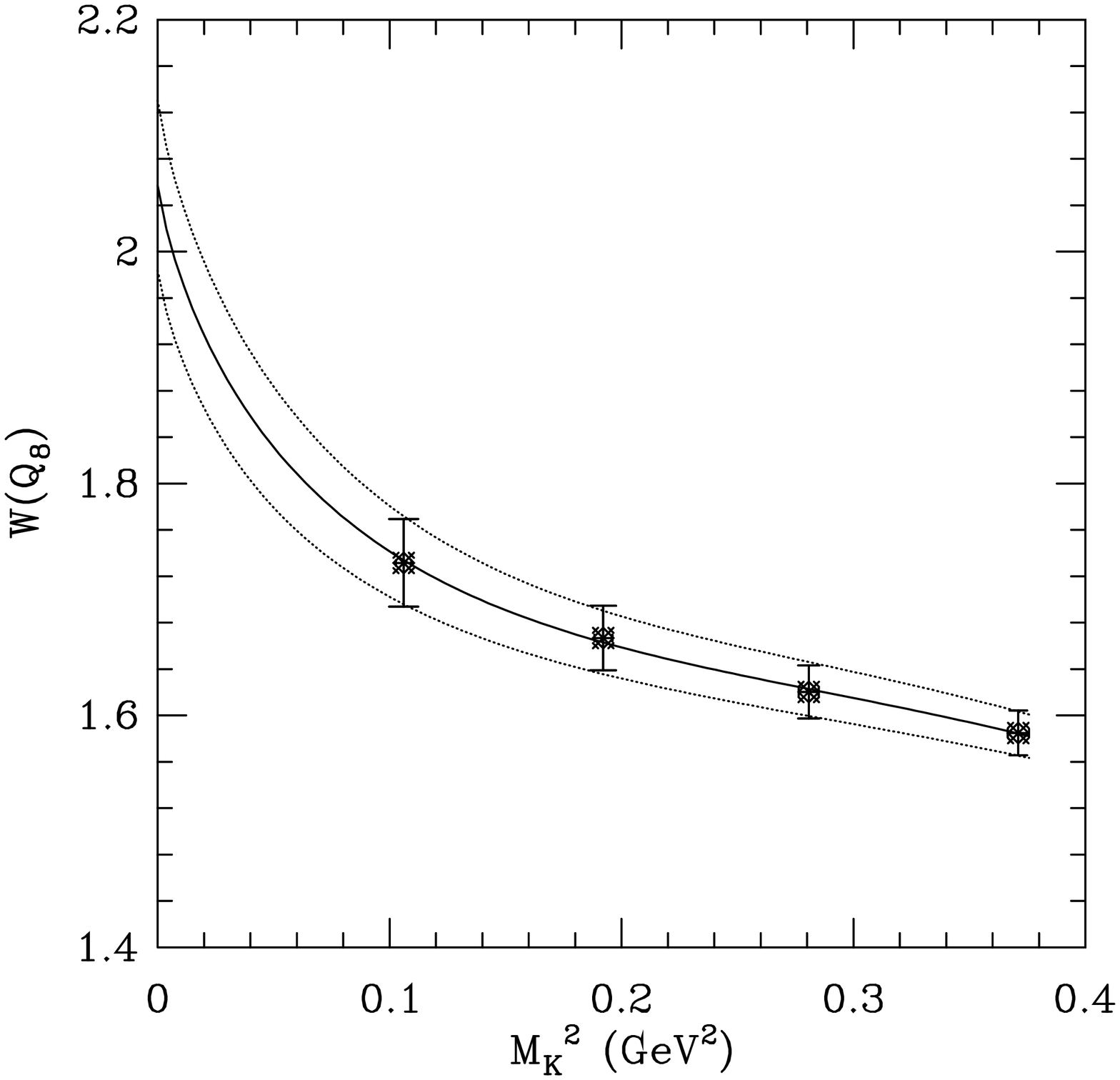, width=0.5\textwidth}
\caption{$W$ for $Q_7$ (left) and $Q_8$ (right)}
\label{fig:WI:Q7-Q8}
\end{figure}
We measure $W$ of Eq.~(\ref{eq:WI:1}) for four different quark masses.
The results for $Q_7$ and $Q_8$ are presented in
Fig.~\ref{fig:WI:Q7-Q8}.
In Fig.~\ref{fig:WI:Q7-Q8} we plot the ratio $W$ as a function of
$m_K^2$.
We fit the data to the following form suggested by quenched chiral
perturbation theory \cite{ref:laiho:1,ref:wlee:1}:
\begin{eqnarray}
f(m_K^2) &=& c_0 \bigg[ 1 + 3 \frac{m_K^2}{(4 \pi f)^2} \log
\bigg( \frac{m_K^2}{(4 \pi f)^2} \bigg)
\bigg]
+ c_1 (m_K^2) + c_2 (m_K^2)^2
\label{eq:WI:fit:chipt}
\end{eqnarray}
Here, even though we know that $c_0=2$ from chiral perturbation
theory, we fit the data by considering $c_0$ a free parameter to see
how well it agrees with the prediction of chiral perturbation theory.
The fitting results are summarized in Tab.~\ref{tab:WI:fit}.
\begin{table}[h!]
\begin{center}
\begin{tabular}{| c | c | c |}
\hline
     & $Q_7$  & $Q_8$ \\
\hline
$c_0$ & 2.1206 $\pm$ 0.0820 & 2.0574 $\pm$ 0.0734 \\
$c_1$ & 4.1970 $\pm$ 0,2298 & 4.6681 $\pm$ 0.2194 \\
$c_2$ & $-$3.9582 $\pm$ 0.4876 & $-$3.9037 $\pm$ 0.4538 \\
\hline
\end{tabular}
\end{center}
\caption{Fit parameters of the Ward identity for $Q_7$ and $Q_8$.}
\label{tab:WI:fit}
\end{table}
Note that the $c_0$ values are consistent with the theoretical
prediction of $c_0=2$ within a statistical uncertainty of
1.5 $\sigma$ for $Q_7$ and 0.8 $\sigma$ for $Q_8$.

Taking into account the fact that we neglect finite volume effects,
higher order corrections in chiral perturbation theory and two-loop
corrections in the matching, the agreement is quite good.
This suggests that we should be able to obtain a solid result for the
$Q_7$ and $Q_8$ matrix elements in the chiral limit using improved
staggered fermions.

\section{Numerical Study on Individual Amplitudes}
\label{sec:num:indivi}
In Ref.~\cite{ref:laiho:1}, Laiho and Soni presented the results of
the chiral log corrections to the individual matrix elements of the
$Q_{(8,8)}$ operator using (partially) quenched chiral perturbation
theory.
For the $\Delta I = 3/2$ amplitude, the chiral behavior is
\begin{eqnarray}
\frac{ \langle \pi^+ | Q_{(8,8)}^{(3/2)} | K^+ \rangle }{f_\pi^2} 
&=& b_0 \bigg[ 1 - 2 \frac{m_K^2}{(4\pi f)^2} 
  \log \Big( \frac{m_K^2}{(4\pi f)^2} \Big) \bigg]
  + b_1 (m_K^2) + b_2 (m_K^2)^2
\label{eq:Q88:3/2:xpt}
\end{eqnarray}
For the $\Delta I = 1/2$ amplitude, the chiral
behavior is
\begin{eqnarray}
\frac{ \langle \pi^+ | Q_{(8,8)}^{(1/2)} | K^+ \rangle_{\rm sub} }
{f_\pi^2} &=& d_0 \bigg[ 1 + \frac{m_K^2}{(4\pi f)^2} \log \Big(
\frac{m_K^2}{(4\pi f)^2} \Big) \bigg] + d_1 (m_K^2) + d_2 (m_K^2)^2
\label{eq:Q88:1/2:xpt}
\end{eqnarray}
where the subscript sub means a kind of subtraction method, 
which will be explained later.
Here, note that the $b_i$ and $d_i$ parameters cannot be determined
from the chiral perturbation theory except for the ratio
$d_0/b_0 = 2$.

In the case of the $ \Delta I = 3/2$ amplitudes, there is no mixing
with unphysical lower dimension operators. 
However, this is not the case for the $ \Delta I = 1/2$ amplitudes.
There is mixing with unphysical lower dimension operators in the next
to leading order ${\cal O}(m_K^2)$.
The mixing coefficients are power divergent of ${\cal O}(1/a^2)$. 
Hence, even though they are suppressed by $m_K^2$, the size of this
mixing dominates the physical signal due to the divergent coefficient
in the range of quark masses that we use in this numerical study.
Therefore it is essential that the mixing with unphysical lower
dimension operators should be subtracted non-perturbatively.
There are many ways to handle this.
The details for different options will be discussed in
Ref.~\cite{ref:wlee:1}.
Here, we focus on the best method that we choose for this numerical
study.
Regardless of the alternative methods to calculate $\langle 0 |
Q_{(8,8)}^{(1/2)} | K^0 \rangle$, if we take the derivative of the
$K^0\rightarrow 0$ amplitude with respect to the strange quark mass
and take the limit of $m_s \rightarrow m_d$, we can select only the
divergent mixing coefficient up to the next leading order of the
quenched chiral perturbation theory.
In other words,
\begin{eqnarray}
\lim_{m_s \rightarrow m_d} \frac{\partial}{\partial m_s} 
\langle 0 | Q_{(8,8)}^{(1/2)} | K^0 \rangle
&=& - \frac{8i}{f} c^r_4 B_0.
\label{eq:Q88:sub}
\end{eqnarray}
where $B_0$ is defined as 
\begin{eqnarray}
m_K^2 = B_0 (m_s + m_d)
\end{eqnarray}
Here, $c^r_4$ is the divergent mixing coefficient in the notation of
Ref.~\cite{ref:laiho:1}.
The details on the derivation of Eq.~(\ref{eq:Q88:sub}) will be
explained in Ref.~\cite{ref:wlee:1}.
At any rate, we use Eq.~(\ref{eq:Q88:sub}) to determine $c^r_4$.
Using this $c^r_4$, we can subtract the divergent mixing away.
Therefore, the final results of the matrix element $\langle \pi^+ |
Q_{(8,8)}^{(1/2)} | K^+ \rangle_{\rm sub}$ do not contain any
contribution from the mixing with unphysical lower dimension
operators.
%

%
%\begin{figure}
%\epsfig{file=Q7I2_chipt.eps, width=0.5\textwidth}
%\epsfig{file=Q7I0_chipt.eps, width=0.48\textwidth}
%\caption{$K^+\rightarrow\pi^+$ amplitudes for $Q_7$ (left: I=2) 
%and (right: I=0)}
%\label{fig:Q7}
%\end{figure}
%

%
\begin{figure}[h!]
\epsfig{file=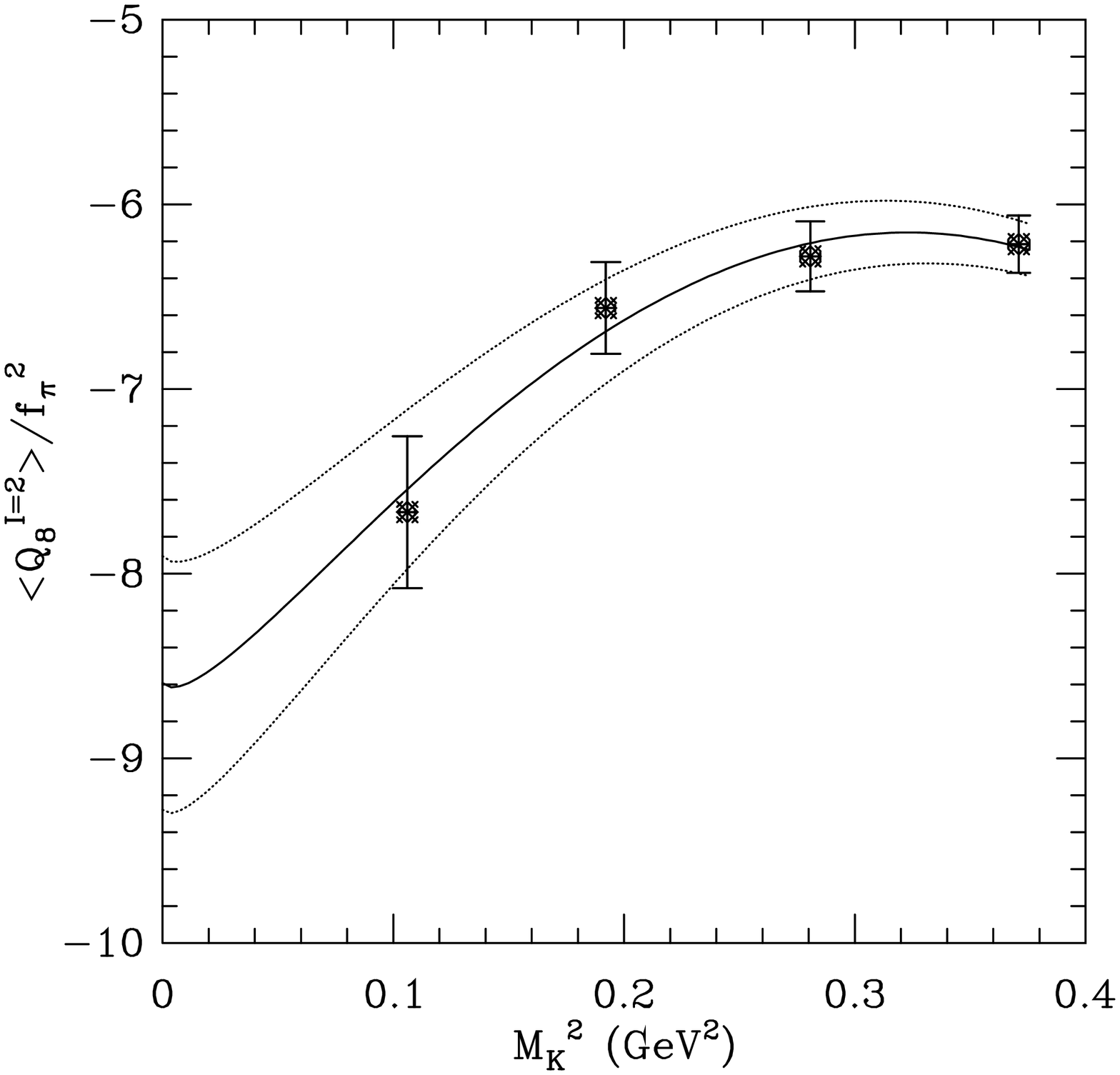, width=0.5\textwidth}
\epsfig{file=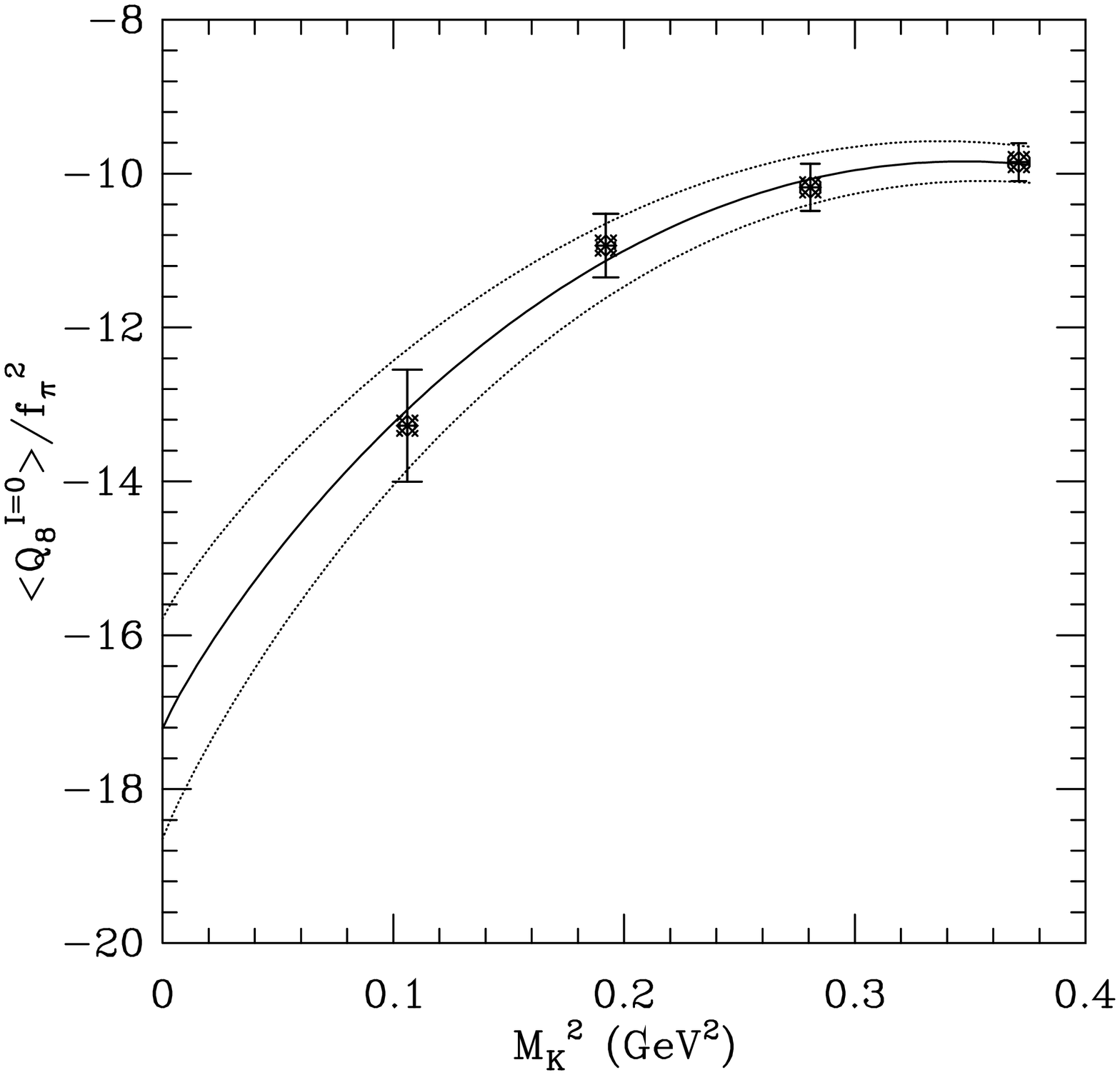, width=0.5\textwidth}
\caption{$\langle \pi^+ | Q_{(8,8)}^{\Delta I} | K^+ \rangle / f_\pi^2
$ matrix elements for $Q_8$ (left: $\Delta I=3/2$) and (right: $\Delta I=1/2$)}
\label{fig:Q8}
\end{figure}

We measure the $K^+\rightarrow\pi^+$ amplitudes of the $Q_8$ operator
for four different quark masses.
The results are presented in Fig.~\ref{fig:Q8}.
Here, we normalize the amplitudes by $f_\pi^2$ for the analysis
convenience without loss of generality.\footnote{
Note that $f_\pi^2$ by itself cannot possess chial logs in the limit
of $m_s = m_d$ in quenched QCD.}
We fit the data to the functional form suggested by the quenched
chiral perturbation theory which is given in
Eq.~(\ref{eq:Q88:3/2:xpt}) and Eq.~(\ref{eq:Q88:1/2:xpt}).
As you can see in Fig.~\ref{fig:Q8}, the intercepts of the $\Delta I = 1/2$
and $\Delta I = 3/2$ amplitudes respect the Ward identity ($d_0/b_0=2$)
within statistical uncertainty.
\section{Gold-plated Ratio}
\label{sec:gold:ratio}
In Ref.~\cite{ref:damir:1}, Becirevic and Villadoro proposed that it
is possible to classify ratios of the $K^+ \rightarrow \pi^+$
amplitudes as gold-plated and silver-plated.
Here, the gold-plated ratio means that it does not contain any chiral
log correction at the one loop level and correspondingly the leading
finite volume effects also cancel off.
This gold-plated ratio ($R$) is defined as
\begin{eqnarray}
R \equiv \frac{\langle \pi^+ | Q_8^{(3/2)} | K^+ \rangle }
{\langle \pi^+ | Q_7^{(3/2)} | K^+ \rangle}
\end{eqnarray}
\begin{figure}[h!]
\begin{center}
\epsfig{file=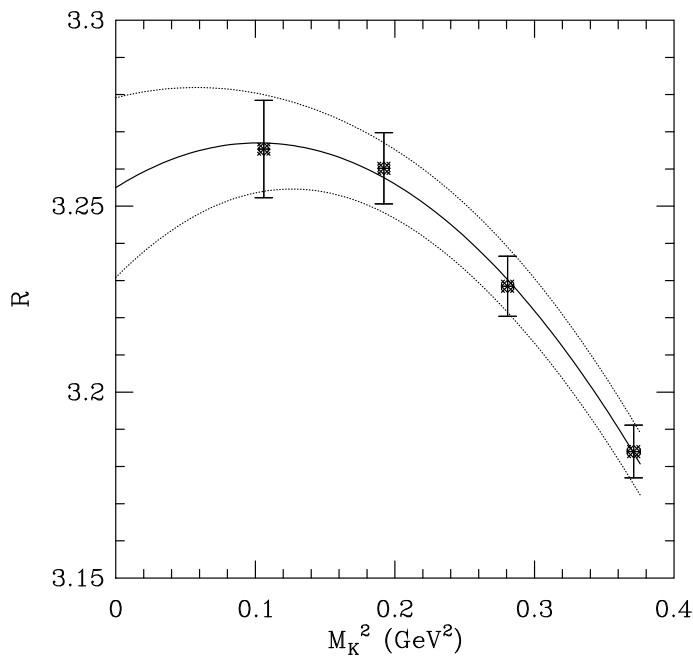, width=0.6\textwidth}
\end{center}
\caption{R vs. $M_K^2$}
\label{fig:R}
\end{figure}
We measure $R$ over four quark masses and the results are presented in
Fig.~\ref{fig:R}.
The quadratic fitting results in Fig.~\ref{fig:R} are
\begin{eqnarray}
R(m_K^2) &=& 3.255(24) + 0.235(154) (m_K^2) - 1.153(247) (m_K^2)^2  
\end{eqnarray}
Here, note that the coefficient of the quadratic term is reasonable
(of ${\cal O}(1)$) whereas the linear term has a relatively tiny
coefficient.
This weak dependence on the linear term might be interpreted as a
signal for the gold-plated ratio.
However, the noticeable dependence on the quadratic term gives us some
doubt on whether we call $R$ gold-plated literally.

\end{document}